\documentclass[article]{aa}
\usepackage{graphicx}
\usepackage{natbib}
\bibpunct{(}{)}{;}{a}{}{,}

\newcommand{\msol}{\hbox{\kern 0.20em $M_\odot$}}
\newcommand{\lsol}{\hbox{\kern 0.20em $L_\odot$}}
\newcommand{\rsol}{\hbox{\kern 0.20em $R_\odot$}}
\newcommand{\sr}{\hbox{\kern 0.20em sr}}
\newcommand{\srmu}{\hbox{\kern 0.20em sr$^{-1}$}}

\newcommand{\g}{\hbox{\kern 0.20em g}}
\newcommand{\gmu}{\hbox{\kern 0.20em g$^{-1}$}}
\newcommand{\kg}{\hbox{\kern 0.20em kg}}
\newcommand{\pc}{\hbox{\kern 0.20em pc}}
\newcommand{\cm}{\hbox{\kern 0.20em cm}}
\newcommand{\m}{\hbox{\kern 0.20em m}}
\newcommand{\km}{\hbox{\kern 0.20em km}}
\newcommand{\nm}{\hbox{\kern 0.20em nm}}

\newcommand{\s}{\hbox{\kern 0.20em s}}
\newcommand{\h}{\hbox{\kern 0.20em h}}
\newcommand{\smu}{\hbox{\kern 0.20em s$^{-1}$}}
\newcommand{\smd}{\hbox{\kern 0.20em s$^{-2}$}}
\newcommand{\an}{\hbox{\kern 0.20em an}}
\newcommand{\anmu}{\hbox{\kern 0.20em an$^{-1}$}}
\newcommand{\yr}{\hbox{\kern 0.20em yr}}
\newcommand{\yrmu}{\hbox{\kern 0.20em yr$^{-1}$}}
\newcommand{\Myr}{\hbox{\kern 0.20em Myr}}
\newcommand{\Mymu}{\hbox{\kern 0.20em Myr$^{-1}$}}
\newcommand{\K}{\hbox{\kern 0.20em K}}
\newcommand{\pcmu}{\hbox{\kern 0.20em pc$^{-1}$}}
\newcommand{\pcmd}{\hbox{\kern 0.20em pc$^{-2}$}}
\newcommand{\pcmt}{\hbox{\kern 0.20em pc$^{-3}$}}
\newcommand{\kms}{\hbox{\kern 0.20em km\kern 0.20em s$^{-1}$}}
\newcommand{\kmpd}{\hbox{\kern 0.20em km$^{2}$}}
\newcommand{\kpc}{\hbox{\kern 0.20em kpc}}
\newcommand{\cms}{\hbox{\kern 0.20em cm\kern 0.20em s$^{-1}$}}
\newcommand{\erg}{\hbox{\kern 0.20em erg}}
\newcommand{\ergs}{\hbox{\kern 0.20em erg}}
\newcommand{\cmpd}{\hbox{\kern 0.20em cm$^2$}}
\newcommand{\cmmd}{\hbox{\kern 0.20em cm$^{-2}$}}
\newcommand{\cmms}{\hbox{\kern 0.20em cm$^{-6}$}}
\newcommand{\cmpt}{\hbox{\kern 0.20em cm$^3$}}
\newcommand{\cmmt}{\hbox{\kern 0.20em cm$^{-3}$}}
\newcommand{\mpd}{\hbox{\kern 0.20em m$^2$}}
\newcommand{\mmd}{\hbox{\kern 0.20em m$^{-2}$}}
\newcommand{\mpt}{\hbox{\kern 0.20em m$^3$}}
\newcommand{\mmt}{\hbox{\kern 0.20em m$^{-3}$}}
\newcommand{\mujy}{\hbox{\kern 0.20em $\mu$Jy}}
\newcommand{\mjy}{\hbox{\kern 0.20em mJy}}
\newcommand{\Mj}{\hbox{\kern 0.20em MJy}}
\newcommand{\jy}{\hbox{\kern 0.20em Jy}}
\newcommand{\ghz}{\hbox{\kern 0.20em GHz}}
\newcommand{\G}{\hbox{\kern 0.20em G}}
\newcommand{\muG}{\hbox{\kern 0.20em $\mu$G}}

\newcommand{\htwos}{\hbox{H$_2$S~$2_{2,0}\rightarrow 2_{1,1}$}}

\newcommand{\sothree}{\hbox{SO~$3_2\rightarrow 2_1$}}
\newcommand{\sosix}{\hbox{SO~$6_5 \rightarrow 5_4$}}
\newcommand{\soiso}{\hbox{$^{34}$SO~$3_2\rightarrow 2_1$}}
\newcommand{\sotwo}{\hbox{SO$_{2}$~$5_{1,5}\rightarrow 4_{0,4}$}}

\begin{document}

\hyphenation{Fe-bru-ary Gra-na-da mo-le-cu-le mo-le-cu-les}

\title{Sulphur-bearing species in the star forming region L1689N}
          \author{V.Wakelam\inst{1},
           A.Castets\inst{1}, C.Ceccarelli\inst{1,2},
          B.Lefloch\inst{2}, E.Caux\inst{3} and L.Pagani\inst{4}}

\institute{
Observatoire de Bordeaux, BP 89, 33270 Floirac, France
\and Laboratoire d'Astrophysique, Observatoire de Grenoble - BP 53,
F-38041 Grenoble cedex 09, France
\and CESR CNRS-UPS, BP 4346, 31028 - Toulouse cedex 04, France
\and LERMA \& FRE 2460 du CNRS, Observatoire de Paris, 61, Av. de
l'Observatoire, 75014 Paris, France}
\offprints{wakelam@observ.u-bordeaux.fr}

\date{Received ; accepted }
\titlerunning{S-molecules in L1689N}
\authorrunning{Wakelam et al.}

\abstract{We report observations of the expected main S-bearing
species (SO, SO$_2$ and H$_2$S) in the low-mass star forming region
L1689N.  We obtained large scale ($\sim 300''$x$200''$) maps of
several transitions from these molecules with the goal to study the
sulphur chemistry, i.e. how the relative abundances change in the
different physical conditions found in L1689N.  We identified eight
interesting regions, where we carried out a quantitative comparative
study: the molecular cloud (as reference position), five shocked
regions caused by the interaction of the molecular outflows with the
cloud, and the two protostars IRAS16293-2422 and 16293E. In the
cloud we carefully computed the gas temperature and density by means
of a non-LTE LVG code, while in other regions we used previous
results. We hence derived the column density of SO, SO$_2$ and
H$_2$S, together with SiO and H$_2$CO - which were observed previously
- and their relevant abundance ratios.  We find that SiO is the
molecule that shows the largest  abundance variations in the shocked
regions, whereas S-bearing molecules show more moderate variations.
Remarkably, the region of the brightest SiO emission in L1689N
is undetected in SO$_2$, H$_2$S and
H$_2$CO and only marginally  detected in SO.  In the other weaker SiO
shocks, SO$_2$ is enhanced with respect to SO.  We propose a schema in
which the different molecular ratios correspond to different ages of
the shocks.  Finally, we find that SO, SO$_2$ and H$_2$S have
significant abundance jumps in the inner hot core of IRAS16293-2422
and discuss the implications of the measured abundances.
\keywords{ISM: abundances -- ISM: molecules -- Stars: formation --
ISM: individual: L1689N, IRAS16293-2422 } 
} 
\maketitle

\section{Introduction}

Low mass star forming regions are composed by at least three main
ingredients: the molecular cloud from which protostars are born, the
protostars themselves, and the shocked regions at the interface
between the cloud and the outflows emanating from the protostars.
These three regions have very different physical conditions, where
temperature, density and also chemical abundances greatly differ
\citep*[e.g.][]{1998ARA&A..36..317V}.  This paper focuses on the
abundance changes occurring to the S-bearing molecules and the
relevant sulphur chemistry.  Depending on the physical condition of
the gas, it is believed that different types of reactions play a role
in the formation of sulphur-bearing molecules.  In molecular clouds,
ion-molecule reactions are the most important
\citep{1974ApJ...187..231O, 1982ApJ...260..590P,1990A&A...231..466M},
whereas in the warm gas of the hot cores and shocks, neutral-neutral
reactions play the major role in forming sulphur species
\citep{1993MNRAS.262..915P, 1997ApJ...481..396C,1998A&A...338..713H,
2001A&A...376L...5K}. Specifically, in warm gas, the abundances of
H$_2$S, SO and SO$_2$ are supposed to increase significantly.  This is
the reason why they are often used to trace shocks
\citep{1993MNRAS.262..915P,1994ApJ...436..741C, 1997ApJ...487L..93B}.
And because of the relatively fast evolution of their chemistry, on
time scale of tens of thousand years, they are good candidates to be
chemical clocks to study the evolution of outflows
\citep{2001A&A...372..899B} and hot cores
\citep{1997ApJ...481..396C,1998A&A...338..713H}.  Overall, it is
widely accepted that in star forming regions the formation of
S-bearing molecules is largely determined during the  cold collapse
phase, when atomic sulphur freezes out on grains and probably forms
H$_2$S. When the protostar starts to heat its environment, H$_2$S
evaporates and it reacts with hydrogen atoms to give sulphur atoms.  S
rapidly reacts with OH and O$_2$ to form SO, that in turn gives SO$_2$
by reacting with OH \citep[e.g.][]{1997ApJ...481..396C}.

In this paper, we present large scale maps of several transitions of
SO, SO$_{2}$ and H$_{2}$S in the molecular cloud L1689N, a molecular
cloud located in the $\rho$ Ophiuchi cloud complex at 120 pc from the
Sun \citep{1998A&A...338..897K}. Based on atomic oxygen observations,
\citet{1999A&A...347L...1C} found that the gas temperature in this
cloud is (26 $\pm$ 0.5)~K and the H$_2$ density is larger than 3
$\times  10^4$ cm$^{-3}$.  L1689N harbors two young protostellar
sources.  The first one is IRAS16293-2422 (hereinafter IRAS16293), a
Class 0 protostar (15 L$_\odot$)  still in the accretion phase
\citep{1986ApJ...309L..47W,1995ApJ...442..685Z,1998ApJ...496..292N,
2000A&A...355.1129C}. Like many other young protostars, IRAS16293 is a
binary system with a total mass around 1.1~M$_\odot$
\citep{2000ApJ...529..477L}, whose two sources are separated by 5$''$,
namely a projected separation of 600 AU.  The structure of the
envelope surrounding IRAS16293 has been reconstructed based on
multifrequency H$_{2}$O, SiO, O and H$_{2}$CO line observations
\citep{2000A&A...355.1129C,2000A&A...357L...9C}. In the outer region
(r $\geq 150$AU), the envelope gas shows molecular abundances typical
of cold molecular clouds. In the inner region (r $\leq 150$AU,
i.e. about 2$''$ in diameter) the abundances of H$_{2}$O, SiO and
H$_{2}$CO jump to abundances typical of the hot cores around massive
protostars.  This structure has been recently confirmed by
\citet{2002A&A...390.1001S}, who modeled the continuum and the line
emission from several other molecules.  The second protostar, 16293E,
is a recently discovered low mass and very young Class 0 source
situated South-East of IRAS16293.  It was detected first by
\citet{1990ApJ...356..184M} as a strong NH$_3$ peak emission. Its
protostellar nature is discussed in \citet[][ hereinafter
CCLCL01]{2001A&A...375...40C}.

\begin{figure}
\centering
\includegraphics[angle=270,width=0.7\columnwidth]{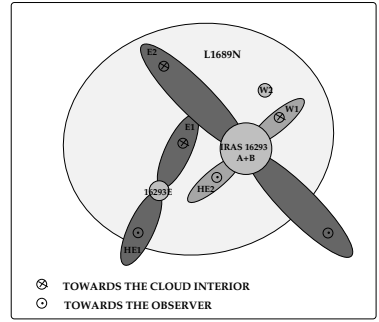}
\caption{Sketch of the region seen face on. The dark grey outflows
have been clearly identified by CCLCL01. Light gray used for the outflow
emanating from IRAS16293 and the W2 source indicates that these
are only assumptions.
}
\label{representation}
\end{figure}

The L1689N region is complex and has long been known to house multiple
outflows
\citep{1986ApJ...311L..85F,1987ApJ...317..220W,1990ApJ...356..184M,
2001ApJ...547..899H}.  The recent work by CCLCL01 claims that the two
protostars IRAS16293 and 16293E drive three bipolar outflows.  Two of
them originate from each of the two components of IRAS16293, while the
third outflow probably emanates from 16293E
(Fig.~\ref{representation}).  In the present article we will adopt the
scheme outlined in CCLCL01 (Fig. 1), but our main conclusions are
substantially unaffected by the actual evolutionary stage (pre-stellar
or  protostellar) of 16293E, questioned in
\citet{2002ApJ...569..322L}.  What is important in the following
discussion is the presence in L1689N of at least one protostar,
IRAS16293, and six regions which shows an enhancement of SiO and/or
H$_2$CO emission, and that are sites of shocked gas marked as  E1, E2,
HE1\footnote{This region is not considered farther in this work
because too weak.}, HE2, W1 and W2 (Fig.~\ref{representation}).  In
particular, CCLCL01 found that the brightest site of SiO emission, E2,
does not show up any H$_2$CO enhanced emission, whereas the brightest
H$_2$CO emission site, E1, is also accompanied by strong SiO emission.
HE2 and HE1 represent a third class, for only H$_2$CO emission is
detected there and no SiO.  The goal of the present work is to study
how the abundances of S-bearing molecules change in all these sites,
compared with the abundances in the IRAS16293 protostar and in the
cloud.

The article is organized as follows.  The observations are presented
in Sect. 2, the results are presented  in Sect. 3, and the column
density determinations are detailed in Sect. 4.  In Sect. 5 we discuss
the results, i.e. the observed changes of SO, SO$_2$ and H$_2$S
abundances with respect to the previously measured SiO and H$_2$CO
abundances and what this may teach us.


\section{Observations}

Observations of large scale maps and specific positions inside the
L1689N  region were performed with the IRAM and SEST telescopes.  We
measured the emission of the following molecules: SO
($3_{2}\rightarrow 2_{1}$ and $6_{5}\rightarrow 5_{4}$ transitions),
SO$_{2}$ ($5_{1,5}\rightarrow 4_{0,4}$ transition)  and H$_2$S
($2_{2,0}\rightarrow 2_{1,1}$ and $1_{1,0} \rightarrow 1_{0,1}$
transitions). We also observed the \soiso line in a few positions in
order to estimate the opacity of the main isotopic line. The
coordinates ($\Delta\alpha, \Delta\delta$) of all maps shown here are
offsets relative to the position of the 16293B component of the binary
system IRAS16293 at $\alpha(2000.0)$ =~16$^{\rm h}$32$^{\rm
m}$22$^{\rm s}$.6, $\delta(2000.0)$ =~-24$^0$28$'$33$''$ \citep[the
16293A component is located 4$^{\prime\prime}$ South and 2$''$ East
from the B component -][]{2000ApJ...529..477L}.

In June 1997 we obtained  a map covering $120\arcsec\times100\arcsec$
in the SO$_2$ molecular line with the IRAM-30m telescope. In November
2001, May and September 2002 we performed additional IRAM observations
of the SO$_2$, $^{34}$SO and H$_2$S $1_{1,0} \rightarrow 1_{0,1}$
lines at some specific positions in L1689N, namely the shocked regions
E1, E2, W1, W2 and HE2 (see Fig. 1 and the Introduction), the two
protostars IRAS16293 and 16293E, and the molecular cloud (at the
position $\Delta\alpha$ = 120\arcsec,  $\Delta\delta$ = 0\arcsec).  In
the following we will refer to these eight positions as the ``key''
positions.  The SEST telescope was used in July 1998 to map an area
covering $300\arcsec \times200\arcsec$ in the SO molecular lines and
to map a smaller area covering $120\arcsec\times100\arcsec$ in the
H$_2$S molecular transition.  Table~\ref{parameters} summarizes the
different observed molecular transitions, together with their
frequencies and upper level energies.

All data were obtained in the position switching mode with an OFF
position located at $\Delta\alpha$~=~-180\arcsec,
$\Delta\delta$~=~0$''$ from the center of IRAS16293. Using the
frequency switching mode we checked  that this position is free of any
C$^{18}$O and $^{13}$CO emission. Even though our observations are not
Nyquist sampled, we smoothed our data to the largest beamsize when
necessary, which is accurate enough, except in the case of
pathological source morphologies. Both with the IRAM and SEST
telescopes, pointing and focus was checked every 2 hours, using
planets, maser sources and strong extragalactic continuum sources. The
pointing corrections were found to be always smaller than 3$''$ and
5$''$ with the IRAM and  SEST telescopes respectively. Polynomial
baselines of order 3 or less have been subtracted from the spectra.
The observing parameters for both telescopes are listed in
Table~\ref{parameters}.  In the following, all intensities will be
given in main-beam brightness temperatures.  Below we give some
information which are specific to the IRAM-30m and SEST observations.

\noindent {\bf IRAM Observations:} The
IRAM 30--meter telescope is located at an altitude of 2920 meters near
the  summit of Pico Veleta in Southern Spain. The \soiso, SO$_2$ and
H$_2$S molecular emissions were observed simultaneously with the 3, 2
and 1mm SIS  receivers respectively, available at the IRAM-30m. The
SO$_2$ emission observed in June 1997 was observed simultaneously
with other molecular transitions not presented here. The image
sideband rejections of all  receivers were always higher than 10 dB.
Typical system temperatures were in the range  T$_{\rm sys} \sim  300
- 600$ K.  The three receivers were  connected to units of the
autocorrelator, set to provide spectral  resolutions of 40, 40, and 80
kHz at 98, 136, and 217 GHz respectively.   At those frequencies, the
velocity resolutions are of the order of 0.1~-~0.2~km~s$^{-1}$.

\smallskip

\noindent {\bf SEST Observations:} The SEST telescope is a 15--meter
single dish millimeter telescope operated jointly by ESO and a
consortium  of Swedish institutions.  It is located at an altitude of
about 2450  meters in La Silla (Chile).  To obtain the SO and H$_2$S
maps shown here, we used the dual 3/1.3mm receiver. System
temperatures  were in the range T$_{\rm sys} \sim 300 - 500$ K. This
receiver was always  connected to the high resolution acousto-optical
spectrometer available at SEST, which provides 2048 channel of 43~kHz
width giving a 86~kHz spectral resolution. At the  frequencies
considered here, that corresponds to a velocity resolution of  0.1 --
0.2 km~s$^{-1}$, comparable to that of the  IRAM data.
\begin{table*}
\begin{tabular}{l|c|c|c|cc|cc|c}
     \hline Transitions & \sothree\ & \sosix\ & \soiso\ &
\multicolumn{2}{|c|}{SO$_{2}$~$5_{1,5}\rightarrow 4_{0,4}$} &
\multicolumn{2}{|c|}{H$_2$S~$2_{2,0}\rightarrow 2_{1,1}$} & H$_2$S
$1_{1,0} \rightarrow 1_{0,1}$ \\ \hline \hline $\nu$ (GHz) & 99.299 &
219.949 & 97.715 & 135.696 & & 216.710 & & 168.762 \\ E$_{up}$/k (K) &
9 & 35 & 9 & 15.6 & & 84 & & 24 \\ n$_{\rm cr}$ (cm$^{-3}$) & $10^{5}$
& $10^{6}$ & $10^{5}$ & $2 \times 10^{6}$ & & $10^{6}$ & & $2 \times
10^{6}$ \\ Telescope & SEST & SEST & IRAM & IRAM & IRAM & SEST & IRAM
& IRAM \\ T$_{sys}$ (K) & 160 & 250 & 140 & 400 & 270 & 220 & 600 &
700 \\ Beam ($''$) & 51 & 24 & 26 & 18 & 18 & 25 & 11 & 14.5 \\ $\rm
B_{eff}$ & 0.75 & 0.5 & 0.75 & 0.59 & 0.71 & 0.5 & 0.57 & 0.65 \\
$\Delta \nu$ (kHz) & 86 & 86 & 40 & 39 & 39 & 86 & 42 & 42 \\
$\Delta\alpha\times\Delta\delta$ ($\arcsec \times \arcsec $) & $280
\times 190$ & $288 \times 216$ & - & 120$\times$100 & - &
120$\times$100 & - & - \\ sampling ($''$) & 24 & 24 & - & 12 & - & 24
& - & - \\ \hline
\end{tabular}
\caption{Observed  molecules together with the observing
parameters. The critical densities (n$_{\rm cr}$) are computed for
temperatures between 50 and 100~K using the collisional coefficients
quoted in Sect.~4. The area mapped ($\Delta\alpha \times
\Delta\delta$)  in each transition and the sampling of the map are
reported in the last two lines. B$_{\rm eff}$ is the beam efficiency
of the telescope.  The symbol ``-'' implies that only observations
towards the ``key'' positions have been obtained.  
}
\label{parameters}
\end{table*}
%

\section{Results}

\begin{table*}
\begin{tabular}{|c|c|c|c|c|c|c|c|c|}    \hline
\multicolumn{2}{|c|}{}

               & SO & SO & $^{34}$SO & SO$_{2}$ & H$_{2}$S &
H$_2$S  & SiO  \\
\multicolumn{2}{|c|}{} & $3_{2} \rightarrow 2_{1}$ & $6_{5} \rightarrow 
               5_{4}$ & $3_{2} \rightarrow 2_{1}$ & 
	       $5_{1,5} \rightarrow 4_{0,4}$ & $2_{2,0} \rightarrow 
               2_{1,1}$ & $1_{1,0} \rightarrow 1_{0,1}$ & $2 \rightarrow 
               1$ \\
              \hline
IRAS   & T$_{MB}$  & 2.9$\pm$0.1 & 3.7$\pm$0.1 & 0.3$\pm$0.1 &
               1.1$\pm$0.1  & 0.5$\pm$0.04 & 6.2$\pm$0.6 & 
               0.35$\pm$0.05 \\
16293-2422 & $\Delta v$  & 2.0$\pm$0.1 & 3.9$\pm$0.0 & 2.8$\pm$0.5 &
               4.2$\pm$0.2 & 5.0$\pm$0.2 & 3.0$\pm$0.0 & 5.0$\pm$0.4 \\
	& v$_{\rm LSR}$ & 4.0 & 3.9 & 3.9 & 3.9 & 3.3 & 3.9 & 4.2 \\
              & $\int T_{MB} \delta v$ & 8.0$\pm$1.2 &
               16.0$\pm$2.5 & 1.0$\pm$0.3 & 4.6$\pm$0.9 & 3.0$\pm$0.5 &
               17.5$\pm$ 3.2 & 1.9$\pm$0.1 \\
\hline
               & T$_{MB}$ & 3.4$\pm$0.1 & 1.2$\pm$0.2 & - & - & $\leq
               0.2$ & 1.1$\pm$0.1 & - \\
16293E & $\Delta v$  & 1.0$\pm$0.0 & 0.8$\pm$0.1 & - &
               - & - & 1.2$\pm$0.1 & - \\
	& v$_{\rm LSR}$ & 3.7 & 3.7 & - & - & - & 3.9 & - \\
              & $\int T_{MB} \delta v$ & 3.8$\pm$0.6 & 1.0$\pm$0.2
               & - & - &$\leq 0.1$ & 0.6$\pm$ 0.1 & - \\
\hline
               & T$_{MB}$  & 2.9$\pm$0.1 & 2.4$\pm$0.2 & 0.5$\pm$0.1 &
               0.7$\pm$0.1 & $\leq 0.2$ & - & 1.1$\pm$0.1 \\
E1       & $\Delta$v & 1.8$\pm$0.2 & 1.6$\pm$0.1 & 1.2$\pm$0.1 &
               3.1$\pm$0.1 & -  & - & 2.8$\pm$0.1 \\
	& v$_{\rm LSR}$ & 4.0 & 3.8 & 3.9 & 4.0 & -  & - & 3.6 \\
              & $\int T_{MB}\delta v$ & 7.3$\pm$1.2 &
               5.6$\pm$1.0 & 0.8$\pm$0.2 & 2.2$\pm$0.4 & $\leq 
               0.3$ & - & 3.3$\pm$0.3 \\
\hline
E2       & T$_{MB}$  & 2.4$\pm$0.2 & 1.1$\pm$0.2 & 0.9$\pm$0.04 &
               $\leq$0.2 & $\leq 0.4$ & - & 0.53$\pm$0.06 \\
LVC      & $\Delta v$  & 1.3$\pm$0.1 &
               0.9$\pm$0.1 & 0.5$\pm$0.02 & - & - & - & 3.2$\pm$0.1 \\
	& v$_{\rm LSR}$ & 3.9 & 3.8 & 3.6 & - & - & - & 4.0 \\
              & $\int  T_{MB}\delta v$ & 3.7$\pm$0.7 & 1.0$\pm$0.2 &
              0.5$\pm$0.1 & $\leq 0.1$ & $\leq 0.4$ & - & 1.8$\pm$0.1 \\
\hline
E2       & T$_{MB}$  & 0.5$\pm$0.15 & $\leq$0.6 &
               $\leq$0.1 & $\leq$0.2 & $\leq$0.4 & - & 0.83$\pm$0.06 \\
HVC      & $\Delta v$ & 6.6$\pm$0.6 & - & - & - & - & - & 6.3$\pm$0.2 \\
	& v$_{\rm LSR}$ & 10.4 & - & - & - & - & - & 11 \\
               & $\int T_{MB} \delta v$ & 3.6$\pm$0.7 & $\leq 0.5$ & $\leq 0.2$
              & $\leq 0.3$ & $\leq 0.8$ & - & 5.6$\pm$0.2 \\
\hline
               & T$_{MB}$  & 2.6$\pm$0.1 & 3.3$\pm$0.2 &
               0.2$\pm$0.04 & 0.3$\pm$0.1 & $\leq$0.2 & - & 
               0.3$\pm$0.1 \\
W1       & $\Delta v$  & 1.8$\pm$0.2 & 1.3$\pm$0.03 & 2.2$\pm$0.2 &
               1.9$\pm$0.3 & - & - & 3.3$\pm$0.8 \\
	& v$_{\rm LSR}$ & 4.0 & 3.3 & 3.7 & 3.3 &  - & - & 4.6 \\
               & $\int  T_{MB} \delta v$ & 5.8$\pm$1.0 & 9.0$\pm$1.5 &
               0.5$\pm$0.1 & 0.7$\pm$0.2 & $\leq 0.4$ & - & 1.1$\pm$0.2\\
\hline
               & T$_{MB}$  & 2.4$\pm$0.2 & 1.6$\pm$0.2 & 0.30$\pm$0.04 &
                 0.7$\pm$0.1 & $\leq$0.2 & - & 0.5$\pm$0.1 \\
W2       & $\Delta v$  & 1.4$\pm$0.1 & 1.6$\pm$0.1 &
               1.1$\pm$0.1 & 1.2$\pm$0.1 & - & - & 2.1$\pm$0.3 \\
	& v$_{\rm LSR}$ & 3.8 & 3.3 & 3.9 & 3.7 & - & - & 3.5 \\
              & $\int  T_{MB} \delta v$ & 3.8$\pm$0.7 & 2.7$\pm$0.5 &
              0.3$\pm$0.1 & 0.9$\pm$0.2 & $\leq 0.3$ & - & 
              1.2$\pm$0.1 \\
\hline
               & T$_{MB}$  & 3.7$\pm$0.1 & 1.4$\pm$0.2 &
               0.6$\pm$0.04 & 0.3$\pm$0.04 & $\leq 0.2$ & - & - \\
HE2      & $\Delta v$ & 0.9$\pm$0.1 & 0.9$\pm$0.1 & 0.9$\pm$0.1 & 1.8$\pm$0.02
               & - & - & - \\
	& v$_{\rm LSR}$ & 3.7 & 3.5 & 3.5 & 3.6 & - & - & - \\
               & $\int  T_{MB} \delta v$ & 5.3$\pm$0.8 & 2.7$\pm$0.5 &
               0.6$\pm$0.1 & 0.6$\pm$0.1 & $\leq 0.3$ & - & - \\
\hline
               & T$_{MB}$  & 3.1$\pm$0.2 & 1.1$\pm$0.2 & 0.6$\pm$0.04 &
               0.2$\pm$0.04 & $\leq$0.3 & 0.8$\pm$0.2 & - \\
Cloud    & $\Delta v$  & 1.0$\pm$0.04 & 0.7$\pm$0.1
               & 0.5$\pm$0.1 &  1.4$\pm$0.2 & - & 1.1$\pm$0.4 & - \\
	& v$_{\rm LSR}$ & 3.8 & 3.8 & 3.8 & 3.9 & - & 3.4 & - \\
              & $\int  T_{MB} \delta v$ & 3.2$\pm$0.6 & 0.8$\pm$0.2 &
0.3$\pm$0.1 &
              0.2$\pm$0.1 & $\leq 0.3$ & 1.0$\pm$0.2 & - \\
\hline
\end{tabular}
\caption{Line parameters ($\rm T_{MB}$ (K), $\Delta v$ (km~s$^{-1}$), 
v$_{\rm LSR}$(km~s$^{-1}$), $\int T_{MB} \delta v$(K~km~s$^{-1}$))
for the various transitions observed at the ``key'' positions: the
two protostars IRAS16293 and 16293E, the shocked regions E1, E2, W1, W2
and HE2, and the reference position in the cloud. The SiO 
observations were previously published in CCLCL01.
When no signal is
detected, we give an upper limit for the intensity equal to 3~RMS
of the relevant spectrum. The parameters are taken from the spectra 
not smoothed.
The symbol ``-'' implies that the relevant transition has not been 
observed or detected.
}
\label{intensity}
\end{table*}
\begin{figure*}
\includegraphics[angle=270,width=15cm]{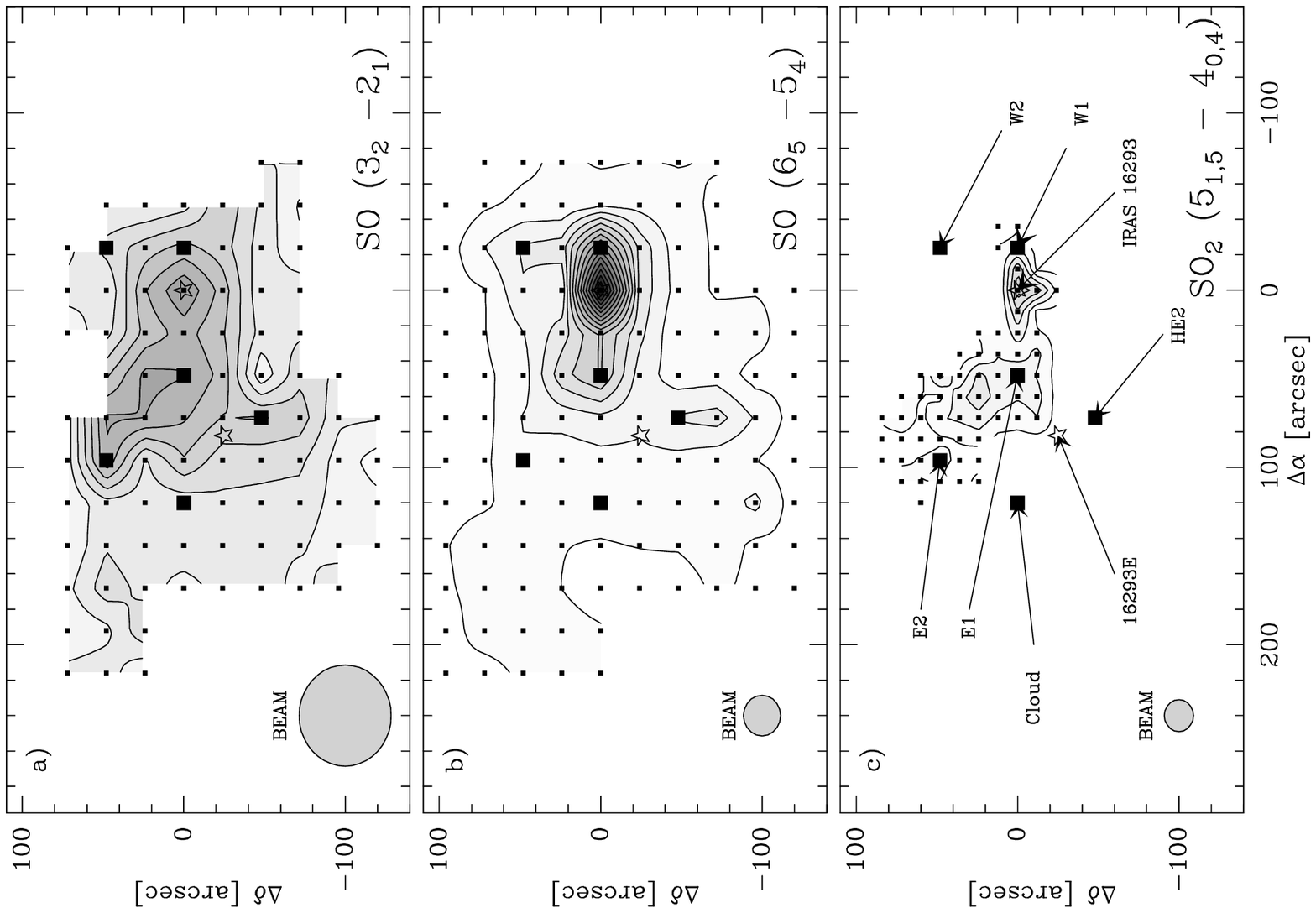}
\caption{Integrated intensity maps of \sothree  (a), \sosix  (b) and
\sotwo  (c). (a) and (b) : first level is 0.5~K~km~s$^{-1}$ with level
step of 1.0~K~km~s$^{-1}$. (c) : first level is 0.3~K~km~s$^{-1}$ with level
step of 1.0~K~km~s$^{-1}$. In each map, the
black points represent the observed positions.  The two stars
symbols show the position of the two protostellar sources. The arrows
point to the eight ``key'' positions, whose spectra are displayed
in Fig.~\ref{Sspect}.
}
\label{3maps}
\end{figure*}
\begin{figure*}
\includegraphics[angle=270,width=18cm]{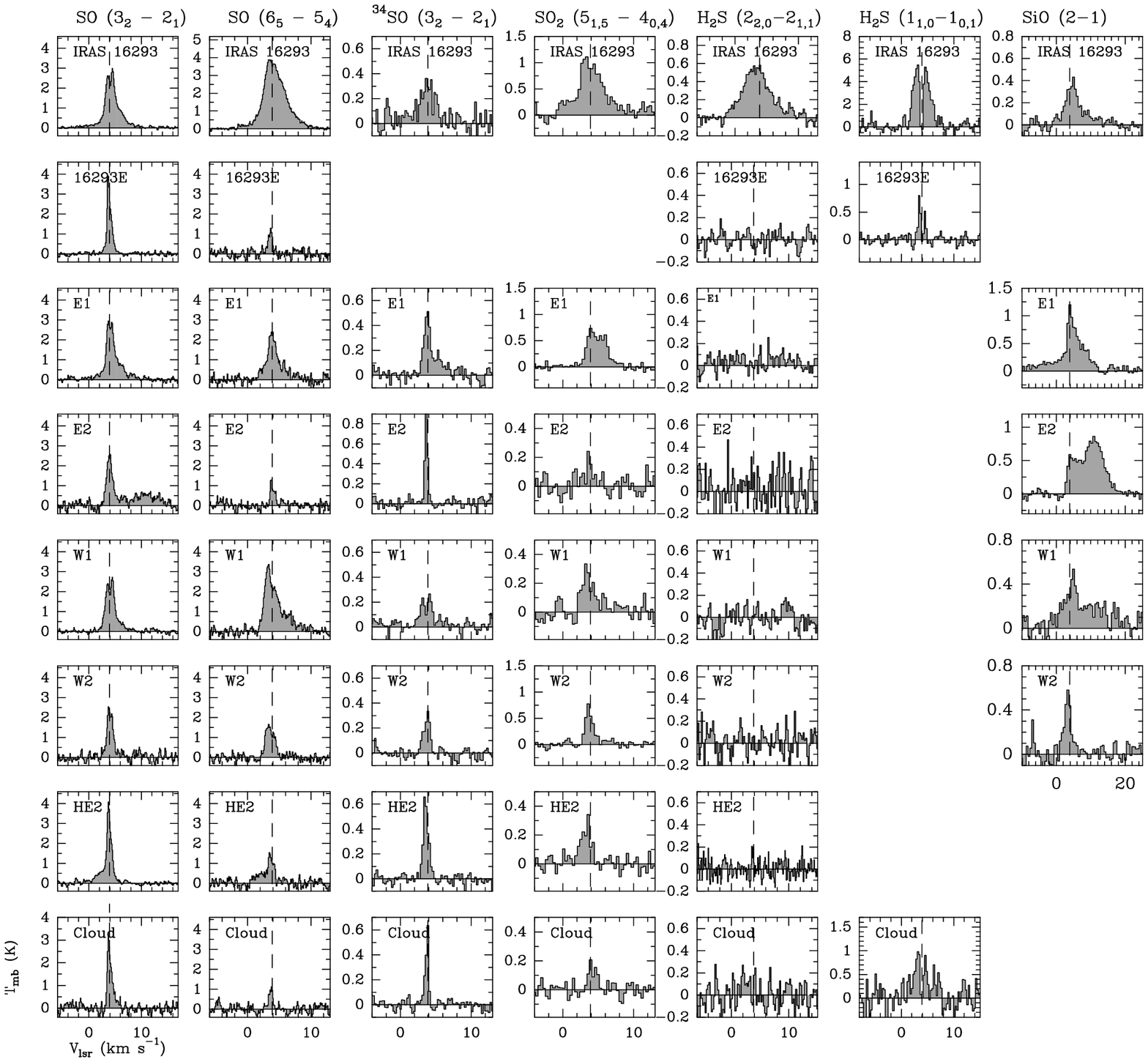}
\caption{Spectra (not smoothed) of \sothree\, \sosix\, \soiso\, \sotwo\,
\htwos\, H$_2$S $1_{1,0} \rightarrow 1_{0,1}$ and SiO $2 \rightarrow 
1$ at ``key'' points (see text) in
the L1689N molecular cloud.
The vertical dashed line on each spectrum shows the position of the cloud
systemic velocity (v$_{LSR}$=3.9~km~s$^{-1}$).
}
\label{Sspect}
\end{figure*}

Figure~\ref{3maps} shows the velocity-integrated intensity maps of the
\sothree, \sosix\ and \sotwo\ lines.  The spectra (not smoothed) of
the six transitions of Table~\ref{parameters} observed towards the
``key''  positions  are displayed in Fig.~\ref{Sspect}, and the
relevant line parameters are reported in Table~\ref{intensity}. 
The SiO $2 \rightarrow 1$ observations, published by CCLCL01, have 
been added on 
Fig.~\ref{Sspect} and in Table~\ref{intensity} for comparisons.

The main-beam brightness peak temperature T$_{\rm MB}$, 
and the linewidth $\Delta$v (FWHM) of Table~\ref{intensity} 
were estimated
from a gaussian fit to the whole line profile, even in the case where 
profiles are double-peaked. 
We also estimated the width of the equivalent gaussian with the same 
integrated and peak intensity as the observed line and found that the 
two methods differ by less than 30\% with the exception of the \sosix\ 
line in W1 which has an equivalent width of 2.5~km~s$^{-1}$ (versus 
1.3~km~s$^{-1}$ in Table~\ref{intensity}).
On the contrary,  the velocity-integrated
intensity is the area integrated in the interval of velocities at the
``zero-intensity'' level. The velocity v$_{\rm LSR}$ is the velocity 
at the peak or at the absorption dip for double-peaked profiles. They are 
roughly the same for each observed transition and position, with 
the exception of E2(HVC), whose emission is red-shifted by $\sim$~6~km~s$^{-1}$.

The first remarkable result is that, against probably naive
expectations, the shocked regions are not necessarily associated with
evident enhancements of the emission of the observed SO, SO$_{2}$ and
H$_{2}$S lines. For example, the E2 region, the brightest SiO 2-1
emission peak (see CCLCL01) is undetected in SO$_{2}$, and only
marginally detected in the lowest SO transition.

The second result is that IRAS16293 is the only site where {\em all}
the molecular lines have been detected, including the high-energy
transition $\htwos$. Quite interestingly, we find that not only the
line profiles are broader than in the cloud but their widths tend to
increase with the upper energy level of the transition.  We will
discuss in the \S 5 the implication of this behavior.

We now discuss in more detail the results toward the eight ``key''
positions.

\subsection{SO emission}

The \sothree\ emission is relatively bright ($\sim 3\K$) all over the
cloud (Figs. \ref{3maps} and \ref{Sspect}).  All the line profiles are
relatively narrow (with linewidths ranging between 1 and $2\kms$) and
peak around the systemic cloud velocity at $\rm v_{lsr}= 3.9\kms$.
The absorption dip around the  systemic velocity seen towards
IRAS16293 is probably due to the absorption by foreground material, as
also seen in the H$_2$CO transitions \citep{2001ApJ...552L.163L}.  On
the other hand, the absorption dip seen in the $^{34}$SO towards W1
may point to a very large SO column density in that direction or two
different kinematic components.  The contrast of the
velocity-integrated SO molecular emission at the various positions in
the cloud is relatively low: the emission peaks towards IRAS16293 but
otherwise it is widespread in the region between W1 and E2
(Fig. \ref{3maps}). Remarkably, there is not a much brighter \sothree\
emission towards the shocked regions with respect to the molecular
cloud.  The only two clear exceptions are represented by E1 and E2,
where a high velocity component appears in the low energy SO
transition (Fig. \ref{Sspect}).  In the following we will analyze
separately the E2 low velocity component centered at $3.9\kms$
(hereinafter E2(LVC)), and the E2 high velocity component at $10\kms$
(hereinafter E2(HVC)), because the two components can be clearly
disentangled.  On the contrary, in E1 the situation is less clear and
hence we didn't pursue a separated analysis of the two components.  We
emphasize that the E2(HVC) component is detected at more than 3 RMS
all along the red lobe of the outflow and it is also observed in the
SiO transitions (CCLCL01).  This high velocity component is not seen
in \sosix\, probably because of unfavorable excitation
conditions. Indeed, we will show in the next section that the gas
density in E2(HVC) is lower than in the ambient gas (see
Table~\ref{ColumnDens}).  This fact, already recognized by CCLCL01,
leads to the supposition that the E2(HVC) may represent the wind
shock, whereas the lower velocity component, E2(LVC), may trace the so
called cloud shock \citep{Hollenbach98IAU}.  Finally, we also see
hints of red and blue wings originating from the molecular outflows
driven either by IRAS16293 or 16293E in the direction of W1, E1 and
HE2 whereas no  wing is seen in W2.

In the higher energy line \sosix\ the emission is peaked towards
IRAS16293 and W1 (cf. Fig.~\ref{3maps}) while the emission from the
cloud is, as expected, weak.  Relatively strong emission is also
detected around E1.  Moreover, the \sosix\ spectra in those positions
have pronounced red and blue wings. The same happens in HE2 where the
\sosix\ spectrum shows clearly the presence of a blue wing.  The
emitting region around IRAS16293 has a characteristic size of $\alpha
\times \delta$ = 39\arcsec\ x 16\arcsec\, namely it is not resolved in
$\delta$.  Finally, the observed SO spectra towards 16293E are similar
to those in the cloud: narrow, intense in the \sothree\ transition and
weak in the \sosix\ transition, and no signs of wings are
evident. Overall, 16293E does not seem to have any enhanced SO
emission compared with the cloud.

\subsection{SO$_{2}$ and H$_{2}$S emission}

Unlike the \sothree\ , the \sotwo\ emission is limited to a few spots
(see Figs.~\ref{3maps} and \ref{Sspect}), specifically toward
IRAS16293, E1 and W2, probably because of the excitation conditions
(the critical density of the \sotwo\ transition is about 20 times the
\sothree\ critical density).  Unfortunately, because of the lack of
other observations around W2 we have no idea of its extent.
Remarkably, no significant SO$_{2}$ emission is detected towards E2. 
SO$_2$ has also been detected at the cloud position. 
The linewidth (1.4 km~s$^{-1}$)
is equivalent to that of N$_2$H$^+$ $1 \rightarrow 0$ -- a good tracer of 
extended, cold and quiescent gas -- observed at the same position (CCLCL01).

Probably because of similar critical densities, the \sotwo\ and
\sosix\ emission shows up in the same regions.  In addition,
with the exception of E1, the $\sotwo$ spectra show characteristics
similar to the \sosix\ spectra, like for example the linewidth, in all
``key'' positions.  On the contrary, in E1 the SO line profiles are
characterized by a strong ambient component with blue- and redshifted
wings of much lower brightness, whereas the high-velocity component
appears as bright as the ambient component in the $\rm SO_2$ line, so
that the whole \sotwo\ linewidth is approximately twice that of the
\sosix\ .

IRAS16293 is the only position where \htwos\ has been detected. The
line is relatively bright and rather broad ($\Delta v =
5.0\kms$). There is no emission detected in shocks nor in the ambient
cloud.  Conversely, the lower H$_{2}$S transition, observed only
toward IRAS16293, 16293E and the cloud, has been detected in all these
positions with a double-peaked profile towards the protostars 
and possibly the cloud position. The absorption dip is likely due to the 
optical depth of the line. In fact, \citet{1991ApJ...366..192M} have 
observed the same H$_2$S transition towards several star forming regions 
and found an opacity of $\sim$10 for similar column densities.

\subsection{Comparison with SiO}

Contrary to the sulphur bearing species, the emission of SiO $2 \rightarrow 1$ 
is stronger in the shocked regions than towards IRAS16293. The spectra in E1 
and W1 show high velocity wings much more marked than for the SO and 
SO$_{2}$ lines. The high velocity component of E2 seen in \sothree\ 
is very strong in SiO $2 \rightarrow 1$ and have similar 
$\Delta v$ ($\sim 6.5~\kms$) 
and v$_{\rm LSR}$($\sim 10~\kms$) for these two transitions.

\section{Column densities}

In this section we estimate the column densities of the observed
species, namely SO, SO$_2$ and H$_2$S, as well of the SiO and H$_2$CO
previously observed by CCLCL01, in the ``key'' positions.  In order to
do that, we first estimate the density and temperature of the gas in
each position.  To derive the gas temperature and density, as well as
the column densities of the different species, we used the theoretical
predictions from an LVG (Large Velocity Gradient) model described in
detail in the next paragraph.  The next three paragraphs describe the
derivation of the gas temperature and densities as well as the column
density of the observed species in the cloud, in the shocked regions,
and in IRAS16293 respectively.  Note that we did not carry out the
analysis on 16293E, because the  observed emission in that position
seems to be dominated by the cloud  emission.

\subsection{LVG model description}

To derive the gas temperature and density, as well as the column
densities, we compared the observed line intensities with the
theoretical predictions from a LVG model, which self-consistently
accounts for the excitation conditions (density and temperature) as
well as the line opacities.  For the escape probability $\beta$ we
used the following function of the line optical depth: $\beta=
\frac{1-{\rm exp[-3\tau]}}{3\tau}$, valid in the case of a homogeneous
and isothermal semi-infinite slab.  More details can be found in
\citet{2002A&A...383..603C}.  Note that the dust emission is neglected
in the present computations.

The spectroscopic data of SO, SO$_2$, H$_2$S, SiO and H$_2$CO are all
taken from the JPL catalogue
\citep[\it{http://spec.jpl.nasa.gov/ftp/pub/catalog/catform.html;
}\rm][]{JPL}.  The collisional coefficients are from
\citet{1994ApJ...434..188G} for SO,  \citet{1987ApJS...64..565P}  for
SO$_{2}$,  \citet{1992ApJ...399..114T} for SiO, and
\citet{1991ApJS...76..979G} for H$_2$CO respectively.  The collisional
coefficients of H$_2$S are not available in literature, and we
estimated them from the collisional coefficients of H$_2$O, multiplied
by a factor 5  \citep[following the discussion
in][]{1996ApJ...468..694T}.  This is a very rough approximation,
giving rise to unfortunately very rough estimates of the H$_2$S column
density.  Furthermore, the collisional coefficients used for the other
molecules are not available in the full range of temperatures probed
by our observations. In particular the low temperature regime is often
missing.  In this case we have extrapolated the coefficients at the
lowest available temperature with a $\sqrt{\rm T_{kin}}$ law.
Finally, our code considers the first 50 rotational levels of each
molecule, unless the relevant collisional coefficients are available
for a lower number of levels.

\begin{table*}
\begin{tabular}{c|cc|ccccc}
\hline
          & Density & T  &
            N$_{\rm SO}$ & N$_{\rm SO_2}$ & N$_{\rm H_2S}$ & N$_{\rm SiO}$ & N$_{\rm H_{2}CO}$ \\
               & ($10^{5} \rm cm^{-3}$) & (K) & ($10^{13} \rm cm^{-2}$) &
               ($10^{13} \rm cm^{-2}$) & ($10^{13} \rm cm^{-2}$) &
               ($10^{13} \rm cm^{-2}$)& ($10^{13} \rm cm^{-2}$) \\
\hline
\hline
IRAS16293 & 250 & 100  & 13000 & 4100 & 4000 & 110 & 750\\
E1      &  1.8 & 150  & 15 & 7.3 & $\leq 96$ & 0.42 & 7.0\\
E2(LVC) & 1.0 & 150 & 3.2 & $\leq 0.44$ & $\leq 260$ & 0.21 & $\leq 0.86$ \\
E2(HVC) & 0.4 & 100 & 7.8 & $\leq 1.1$ & $\leq 1200$ & 1.0 & $\leq 1.8$ \\
W1      & 2.6 & 150  & 24 & 2.4 & $\leq 72$ & 0.15 & 3.4\\
W2      & 1.8 & 80  & 8.7 & 2.9 & $\leq 76$ & 0.15 & 3.6 \\
Cloud     & 0.3 & 26  & 30 - 80 & 2 & 1.5 & 0.02 & 2 \\
\hline
\end{tabular}
\caption{Gas temperature and density, and column densities of 
 SO, SO$_2$, H$_2$S, SiO and $\rm H_{2}CO$ in six ``key'' positions.
The column densities are beam-averaged in the shocked regions (on 27$''$ 
for SiO, 26$''$ for H$_{2}$CO, 24$''$ for SO, 18$''$ for SO$_{2}$ and 
25$''$ for H$_{2}$S),
whereas they are corrected from the beam dilution in IRAS16293.
Errors on the estimates are around 15\% if considering only the statistical
uncertainties. However, a factor two of uncertainty has to be considered,
when the uncertainty on the density and temperature of the emitting
gas is considered.
}
\label{ColumnDens}
\end{table*}
%
%
%
\subsection{Cloud}

\begin{figure}
\centering
\includegraphics[angle=90,width=9cm]{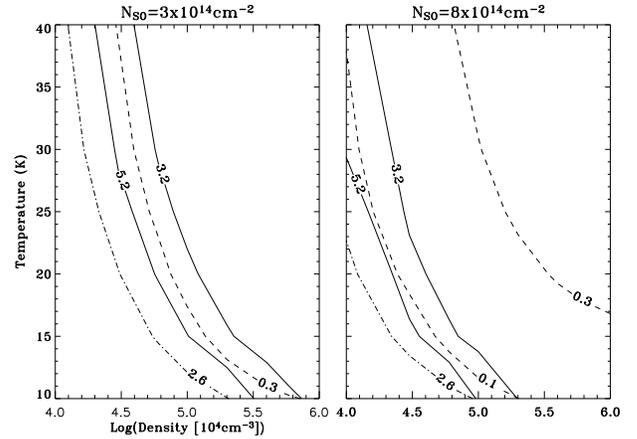}
\caption{Cloud reference position modeling.
Theoretical predictions as function of the gas temperature and density for
two different SO column densities. The curves refer the values of  
three observed quantities taking into account the uncertainties.
Solid lines: observed integrated intensity ratio
\sothree\ over \sosix\ , between 3.2 and 5.2.
Dashed lines: escape probability, derived from the $^{34}$SO and $^{32}$SO
$3_{2}\rightarrow 2_{1}$ line ratio, between  0.1 and 0.3.
Dot-dashed line: the minimum integrated intensity
of the \sothree\ transition, 2.6~K~km~s$^{-1}$.
Note that the
intensity increases with increasing temperature, i.e. towards the upper
part of the panel. Analogously, the lower limit of the escape
probability (0.1) in the left panel is not shown because it lies
outside the plot.
}
\label{ratio}
\end{figure}

To estimate the gas density and temperature, and the SO column density
in the cloud reference position we simultaneously best-fitted the
$^{34}$SO and $^{32}$SO $3_{2}\rightarrow 2_{1}$ observations (that
give a line opacity for $^{32}$SO $\tau \sim$ 1.5), and the $^{32}$SO
$3_{2}\rightarrow 2_{1}$ and $6_{5}\rightarrow 5_{4}$ observations
 (whose ratio, together with the  $^{32}$SO $3_{2}\rightarrow
2_{1}$ intensity, gives pairs of gas temperature and density
values). Note that we used the elemental ratio
$^{34}$S/$^{32}$S = 22.5
\citep{1994ARA&A..32..191W,1996A&A...305..960C,1998A&A...337..246L},
and we assumed that the emission fills up the beam,  when comparing
observations  obtained with different telescopes.  The result of the
modeling is shown in Fig. \ref{ratio}.  We found that the ensemble of
our observations constrains the SO column density in the cloud to be
in the range 3 to 8 $\times 10^{14}$ cm$^{-2}$.  Adopting the gas
temperature of 26 K, as found by \citet{1999A&A...347L...1C}, we
derive a density of  $3 \times 10^{4} \rm cm^{-3}$.

The column densities of the other species, namely SO$_2$ and H$_{2}$S,
have been derived assuming the same density and temperature. The SO$_2$ 
abundance was found to be $4\times 10^{-8}$, a value equivalent to the ones 
obtained in other clouds (see Table~\ref{clouds}). In the
case of H$_{2}$S we cross-checked the column density derivation with
LTE computations, because of the rough approximation of the
collisional coefficients. For this we assumed an excitation
temperature of 5~K, as  found by \citet{1989ApJ...345..828S} and
\citet{1995PASJ...47..845H} in similar molecular clouds (L134N and
TMC-1) for several molecules whose fundamental transitions have
A-coefficients similar to that for the H$_{2}$S $1_{1,0} \rightarrow
1_{0,1}$ transition. In fact \citet{1989ApJ...345L..63M} used an
excitation temperature of 5~K for the H$_{2}$S $1_{1,0} \rightarrow
1_{0,1}$ transition to compute the column density of H$_{2}$S in the
molecular clouds L134N and TMC-1, which have temperatures and
densities similar to L1689N.  Note that increasing the excitation
temperature by a factor two, would decrease the H$_{2}$S column
density by the same amount. We assumed a  H$_{2}$S ortho-to-para ratio
equal to 3. Finally the SiO and H$_{2}$CO column densities are  taken
from the modeling by \citet{2000A&A...357L...9C} and
\citet{2001A&A...372..998C} respectively \footnote{Although no 
SiO emission was detected towards the cloud position, 
\citet{2000A&A...357L...9C} modeled
the observed SiO (from J$_{up}$ = 1 to 8) emission towards
IRAS16293, taking into account the envelope physical structure,
and found that the SiO abundance is 4.0$\times$10$^{-12}$ in the 
outer envelope. In analogy with other molecules for which the abundances 
found in the outer envelope are equal to
the (measured) abundances in the surrounding molecular clouds, we 
assumed 4.0$\times$10$^{-12}$ for the SiO abundance in L1689N. 
In a similar way, although weak formaldehyde emission is detected 
at the cloud position, we used the abundance found in the 
IRAS16293 outer envelope for the cloud too. }.  The column densities
derived for all species are reported in Table \ref{ColumnDens} and the
abundances in Table~\ref{clouds}.
\begin{table}
\begin{tabular}{c|ccccc}
\hline 
Cloud   & x(SO) & x(SO$_2$) & x(H$_2$S) & x(SiO) & x($\rm
H_{2}CO$) \\
& ($10^{-9}$) & ($10^{-9}$) & ($10^{-9}$) & ($10^{-12}$)
& ($10^{-8}$) \\
\hline
\hline
L1689N & 6 - 16 & 0.4 & 0.3 & 4 & 0.04 \\
L134N & 0.6 - 10 & 0.3 - 2.5 & 3 &$\leq 3.6$ & 2 \\
TMC-1 & 0.3 - 4 & 2 - 6 & 0.7 & $\leq 2.4$ & 7 \\
\hline
\end{tabular}
\caption{
Abundances in molecular clouds.  References: L1689N: SO, SO$_{2}$ and
H$_{2}$S from this work, SiO from \citet{2000A&A...357L...9C} and
H$_{2}$CO from \citet{2001A&A...372..998C}; L134N:
SO and SO$_2$ abundances from \citet{1989ApJ...345..828S}, H$_{2}$S
from \citet{1989ApJ...345L..63M}, SiO from
\citet{1989ApJ...343..201Z}, $\rm H_{2}CO$ from
\citet{1992IAUS..150..171O}; TCM-1: SO abundances from
\citet{1995PASJ...47..845H} and \citet{1997ApJ...486..862P}, SO$_2$
from \citet{1983A&A...127L..10I}, H$_{2}$S from
\citet{1989ApJ...345L..63M}, SiO from \citet{1989ApJ...343..201Z},
$\rm H_{2}CO$ from \citet{1998cpmg.conf..205O}.}

\label{clouds}
\end{table}
\subsection{Shocked regions}

We have used the gas temperature and density derived by
CCLCL01 in E1, E2(LVC), E2(HVC), W1,
W2\footnote{Since no SiO emission is detected in HE2, the density and
temperature cannot be constrained in this position.}, i.e.  in the
shocked regions,  by comparing  the SiO observed lines (J from 2 to 5)
with the relevant LVG model.  The uncertainty in the derived densities
and temperatures is around a factor two.  We did not try to derive
independently the gas temperature and density for each molecule,
because of the lack of enough usable transitions. In fact, of the two
SO transitions, the lowest lying transition is dominated by the cloud
emission so it cannot be used to probe the shocked gas, and only one
transition of SO$_2$, H$_{2}$S and H$_{2}$CO respectively has been
detected.  We therefore derived the column densities of those last
four species assuming the same density and temperature derived from
the SiO observations.  This is an approximation  that does not take
into account the possible structure of the shocked gas, but the
results are indeed not much affected by this assumption.  In principle
each molecular species may originate in slightly different physical
conditions and hence the computed column density may be consequently
mis-evaluated. In practice, though, the error associated with this
approximation is lower than about a factor two.  
For example, Lis et al. (2002), using H$_2$CO transitions,  
found different values for the temperature
(45 K) and density ($2 \times 10^6$ cm$^{-3}$) in E1. 
Even taking those values, the derived column density
of the four molecules listed in Table \ref{ColumnDens} wouldn't change
by more than a factor two with respect to those quoted in the table.
Finally, we also assumed that all the used lines are
optically thin \citep[e.g. ][]{1994ApJ...428..680B}.  The column
densities, averaged on the relevant beam, derived for all species in
each of the four shocked sites are reported in Table~\ref{ColumnDens}.
Unfortunately, since we don't have observations of the H$_{2}$S lowest
transitions in the shocks, the column density of this molecule is very
poorly constrained.

\subsection{IRAS16293}

As mentioned in the Introduction, the envelope of IRAS16293 is formed
by (at least) two components: an inner core characterized by a high
temperature and an outer cold envelope with a temperature close to the
cloud temperature. The hot core is a small region ($\sim 2''$ in
diameter) whose temperature is about 100~K and the density $2.5\times
10^{7} \cmmt$ \citep{2000A&A...355.1129C,2002A&A...390.1001S}.  At
that temperature the grain  mantles evaporate injecting into the gas
phase their constituents
\citep{2000A&A...355.1129C,2000A&A...357L...9C}. The outer envelope is
more extended ($\sim 20''$) and colder, with a temperature of 30~K (a
little larger than the L1689N temperature) and a density of $2.5\times
10^{5} \cmmt$ \citep{2000A&A...355.1129C,2002A&A...390.1001S}.  The
molecular abundances in the outer envelope are similar to those in the
molecular cloud, with the possible exception of the deuterated
molecules. On the contrary, in the inner core, several molecules,
those believed to be released from the grain mantles or formed from
those evaporated molecules, undergo a jump in their abundances.  In
the following we focus on the abundance of the S-bearing molecules in
the inner hot core. \citet{1994ApJ...428..680B}, using multifrequency 
observations, constructed the rotational
diagrams of SO and SO$_2$ and derived rotational temperatures 
relatively large ($\sim$80~K). It is therefore very likely that the
bulk of the emission for both molecules originates in the inner core. 
 Furthermore, all linewidths are larger than 3~km~s$^{-1}$ and increase with 
the upper level energy of the transition, once again 
arguing for an inner core origin.
Pursuing this hypothesis we computed the column densities using these 
rotational diagrams, corrected
for the beam dilution, reported in Table~\ref{ColumnDens}.  In
estimating the H$_{2}$S column density we assumed that the
$2_{2,0}\rightarrow 2_{1,1}$ line emission originates entirely in the
hot core, as it is suggested by its observed relatively large
linewidth.  Using then the estimates of the H$_2$ column density in
\citet{2000A&A...355.1129C} ($7.5 \times 10^{22} \rm cm^{-2}$) we
derive the abundances reported in Table~\ref{coresabund}. Note that,
provided that we use our volume density in the \citet{2002A&A...390.1001S}
model, our derived abundance compare extremely well with their 
abundances (which, as a result, is multiplied by a factor five), 
supporting the validity of our method.
Actually \citeauthor{2002A&A...390.1001S} found
abundances a factor 5 lower than ours, because they used a factor 
five larger density than us in the
inner core. The two density estimates differ because of the different
diagnostics used to derive them: \citeauthor{2002A&A...390.1001S} used
the continuum spectral energy distribution, whereas
\citeauthor{2000A&A...355.1129C} used the water line spectrum.  When
CO lines are used instead, \citeauthor{2002A&A...390.1001S} found the
same density in the inner region (Sch\"{o}ier private communication).
We use therefore the estimate by \citet{2000A&A...355.1129C}, but keep
in mind that we may be overestimating the abundances by a factor 5.
Finally, the SiO and $\rm H_{2}CO$ column densities are taken from the
\citet{2000A&A...355.1129C,2000A&A...357L...9C,2001A&A...372..998C}
modeling.

\begin{table*}
\begin{tabular}{lccccc}
\hline
     & SO/H$_2$ & SO$_{2}$/H$_2$ & H$_{2}$S/H$_2$ & SiO/H$_2$ &
H$_{2}$CO/H$_2$ \\
\hline
\hline
IRAS16293 & $1.7 \times 10^{-6}$ & $5.4 \times 10^{-7}$ & $5.3 \times
10^{-7}$ & $1.5 \times 10^{-8}$ & $1.0 \times 10^{-7}$ \\
L1157-mm & $5.0 \times 10^{-9}$ & $3.0 \times 10^{-8}$ & $1.1 \times
10^{-8}$ & - & $4-6 \times 10^{-9}$ \\
Orion-KL & $1.5 \times 10^{-7}$ & $9.4 \times 10^{-8}$ & $5.0 \times
10^{-6}$ & $6.0 \times 10^{-9}$ & $7.0 \times 10^{-9}$  \\
G10.47 & $3.0 \times 10^{-9}$ & $1.8 \times 10^{-8}$ & $\ge 4.0 \times
10^{-9}$ & - & - \\
G29.96 & $\ge 5.0 \times 10^{-9}$ & $2.0 \times 10^{-8}$ & $\ge 1.5 \times
10^{-8}$ & - & - \\
G75.78 & $3.0 \times 10^{-9}$ & $\ge 2.0 \times 10^{-9}$ & $\ge 1.0 \times
10^{-9}$ & - & - \\
G9.62 & $\ge 4.0 \times 10^{-9}$ & $1.0 \times 10^{-8}$ & $\ge 8.0 \times
10^{-9}$ & - & - \\
G12.21 & $\ge 8.0 \times 10^{-10}$ & $\ge 5.0 \times 10^{-10}$ & - & - & - \\
G31.41 & $4.0 \times 10^{-9}$ & $1.2 \times 10^{-8}$ & $\ge 3.0 \times
10^{-9}$ & - & - \\
G34.26 & $6.0 \times 10^{-9}$ & $1.5 \times 10^{-8}$ & $\ge 2.0 \times
10^{-9}$ & - & - \\
BF-Class 0 & $3.0 \times 10^{-9}$ & $4.0 \times 
10^{-10}$ & $2.0 \times 
10^{-9}$ &  - &  - \\ 
\hline
\end{tabular}
\caption{Abundances of SO, SO$_2$, H$_2$S, SiO and H$_{2}$CO with
respect to H$_{2}$ in the hot core of IRAS16293.
The abundances of SO, SO$_2$ and H$_2$S are estimated by the present work
observations, assuming a H$_2$ column density of
$7.5 \times 10^{22} \rm cm^{-2}$ \citep{2000A&A...355.1129C}.
The SiO and H$_{2}$CO abundances are taken from \citet{2000A&A...355.1129C}
and \citet{2000A&A...357L...9C} respectively.
References of the other sources:
L1157-mm(IRAS 20386+6751): \citet{1997ApJ...487L..93B};
Orion-KL: SO, SO$_{2}$, SiO and H$_{2}$CO from \citet{1995ApJS...97..455S},
H$_{2}$S from \citet{1990ApJ...360..136M};
G10.47, G29.96, G75.78, G9.62, G12.21, G31.41 and G34.26:
\citet{1998A&A...338..713H}; BF-Class 0: 
\citet{2003A&A...399..567B}.
}
\label{coresabund}
\end{table*}
%

\section{Discussion}

The abundance ratios between the observed species are reported in
Table~\ref{clouds}, \ref{coresabund} and \ref{AbuRatio}.  They show
variations up to two orders of magnitude.  In the following we analyze
in detail the variations associated with the cloud, the shocked
regions and the protostar IRAS16293 respectively.

\subsection{The cloud}

Adopting the H$_2$ column density derived by
\citet{1999A&A...347L...1C} by means of atomic oxygen observations 
in L1689N, $5\times 10^{22}$ cm$^{-2}$,
and using the column densities derived in the previous section
(Tab.~\ref{ColumnDens}) we obtain the abundances reported in
Tab.~\ref{clouds}. In the table we also report, for comparison, the
abundances measured in L134N and TMC-1, two among the best studied
molecular clouds.  With the possible exception of formaldehyde, which
seems to be underabundant, L1689N has abundances typical of other cold
molecular clouds and it is therefore interesting to study how those
abundances change in the region either because of the presence of
shocks or the presence of a protostar (IRAS16293).

\subsection{Shocked regions}
\begin{table*}
\begin{tabular}{c|ccccccc}
\hline Region & $\rm SO/H_{2}CO$ & SO$_2$/SO & $\rm SO_{2}/H_{2}CO$ & H$_2$S/SO & SiO/SO & SiO/SO$_{2}$ & $\rm SiO/H_{2}CO$ \\
\hline
\hline
Cloud        & 28 & $\sim 0.04$ & 1 & $\sim 0.03$ & $\sim 4 \times 10^{-4}$ & 
0.01 & 0.01 \\
\hline
E1          & 2 & 0.5 & 1.1 & $\leq 6$ & 0.03 & 0.06 & 0.06 \\
E2(LVC)      & $\geq4$ & $\leq 0.1$ & - & $\leq 80 $ & 0.06 & $\geq0.5$ & $\geq0.2$  \\
E2(HVC)     & $\geq 5$ & $\leq$0.1 & - & $\leq 150 $ & 0.1 & $\geq0.9$ &
$\geq0.5$ \\
W1          & 7 & 0.1 & 0.7 & $\leq 3$ & $6\times 10^{-3}$ & 0.06 & 0.04 \\
W2          & 2.5  & 0.3 & 0.8 & $\leq 9$ & 0.02 & 0.05 & 0.04 \\
\hline
\hline
L1157-B1 & 0.6 - 1 & 0.6 - 1 & 0.4 - 1 & 0.8 - 1.3 & 0.2 - 0.3 & 0.3 &
0.1 - 0.3 \\
L1157-B2 & 1 & 1 - 3 & 1 - 3 & 0.7 - 2 & 0.1 - 0.4 & 0.1 & 0.1 - 0.4 \\
CB3 & - & $\sim$1 & - & $\leq$1 & $\ll$1 &  $\ll$1 & -\\
L1448 & - & - & - & - & 0.3 & - & - \\ 
B1 & - & - &  - & - & 0.1 & - & - \\ 
CepA & - & - &  - & - & 0.08 & - & - \\ 
NGC 2071 & - & - &  - & - & 0.01 - 0.05 & - & - \\
\hline
\end{tabular}
\caption{Abundance ratios in the shocked regions and in the cloud,
as derived from the column densities of Tab.~\ref{ColumnDens}.
References for other outflows: L1157-B1 and
L1157-B2: \citet{1997ApJ...487L..93B}; CB3: \citet{1999A&A...350..659C};
L1448, B1 and CepA:
\citet{1992A&A...254..315M}; NGC 2071: \citet{1993ApJ...403L..21C}.}
         \label{AbuRatio}
\end{table*}
\begin{figure*}
\centering \includegraphics[angle=90,width=14cm]{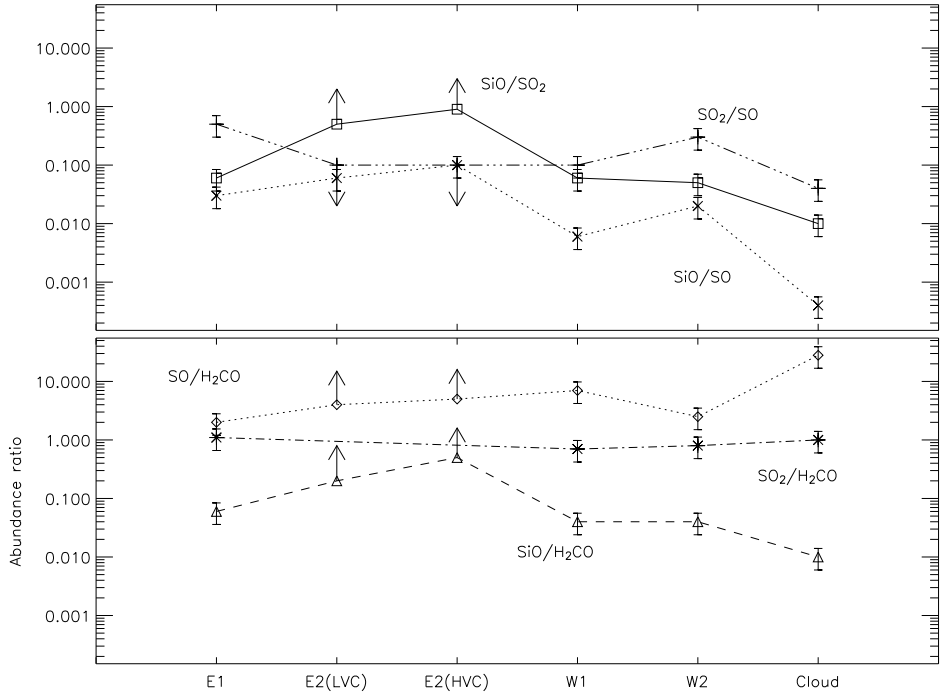}
\caption{The abundance ratios observed in the different shocked
regions and in the molecular cloud and presented in
Table~\ref{AbuRatio}: SO/H$_{2}$CO, SO$_{2}$/H$_{2}$CO and
SiO/H$_{2}$CO on the lower panel and SO$_{2}$/SO, SiO/SO$_{2}$ and
SiO/SO on the upper panel. The arrows represent upper and lower 
limits of the SO$_{2}$/SO, SiO/SO$_{2}$,
SO/H$_{2}$CO and SiO/H$_{2}$CO ratios respectively.}
\label{Graphabund}
\end{figure*}

In the shocked regions, with the exception of E2, it is practically
impossible to derive the absolute abundances of the observed
molecules, because of the difficulty to derive reliable estimates of
the H$_2$ column densities.  In fact, usually the H$_2$ column
densities are estimated from the CO millimeter observations converting
the measured CO column densities into H$_{2}$ column densities.  The
method relies on the capacity to disentangle the contribution of the
cloud from the shocked gas in the CO emission.  However, this is
feasible when the shocked gas emits at relatively large velocities,
i.e. when the two contributions can be separated based on their
spectral properties \citep[see for example ][]{1997ApJ...487L..93B}.
In the specific case of the outflows in L1689N, the projected velocity
of the shocks is too small and the cloud and shocked gas cannot be
disentangled, except in the E2 position, where a high velocity
component is  present.  In order therefore to estimate the H$_2$
column density in the shocked regions we would need high spatial
resolution observations, because single dish CO measurement are totally
dominated by the cloud emission.  For this practical reason, we
computed only abundances ratios, as summarized in
Table~\ref{AbuRatio}.  Note that the quoted ratios have been obtained
dividing column densities averaged on different beam sizes ($18''$ for
SO$_{2}$ observations and $\sim 25''$ for the others), so that the
SO$_{2}$ ratios may be off by a factor two (in the shocked regions).
Fig.~\ref{Graphabund} shows a graphic representation of the derived
abundances ratios in the shocked regions and in the cloud.

The first result to note is that, with the exception of the
SO$_{2}$/H$_{2}$CO ratio, all other ratios in the shocked regions
differ by more than a factor two when compared to the ratios in the
cloud.  In particular, the largest variations, by up a factor 100, are
shown by the ratios with the SiO, confirming that SiO is the best
molecule to trace shocks
\citep{1997A&A...322..296C,1997A&A...321..293S,2002ApJ...567..980G}.

A second robust result is that SO$_2$ is enhanced with respect to SO
in the E1, W1 and W2 shocks, when compared to the cloud  (by up a
factor 10).  This also confirms a rather general theoretical
prediction that SO$_{2}$ is overproduced with respect to SO in the
shocked gas, because of endothermic reactions that lock most of the
gaseous sulphur in SO$_2$ on timescales relatively small $\sim
10^3$ yr \citep[e.g.][]{1993MNRAS.262..915P}.  Remarkably, SO$_2$/SO is
lowest in the strongest SiO shock of the region, E2.  CCLCL01 argued that
E2 is an older shock based on H$_{2}$CO and SiO observations compared 
with models predictions and this would agree with the low SO$_2$/SO
ratio, as SO$_2$ is expected to be converted into atomic sulphur at
late stages $\geq 3 \times 10^4$ yr \citep[e.g.][]{1998A&A...338..713H}.

The third result, is the constant SO$_{2}$/H$_{2}$CO ratio, within a
factor 2, in the shocked regions E1, W1 and W2, and in the cloud.  Why
SO$_{2}$/H$_{2}$CO has the same value in these shocked regions and in
the cloud seems a remarkable coincidence for which we do not have an
explanation.  On the contrary, the constant SO$_{2}$/H$_{2}$CO ratio
in these shocked regions may be due to the timescales involved in the
formation and destruction of the two molecules, even though the
mechanisms of SO$_2$ and H$_2$CO formation in shocks are expected to
be different. On the one hand, formaldehyde is thought to be formed
onto the grain mantles,  released into the gas phase because of the
mantle sputtering in the shock, and finally converted into more
complex molecules by gas phase reactions on timescales larger than
$\sim 3 \times 10^4$ yr \citep[][]{1992ApJ...399L..71C}.  SO$_2$ follows a
totally different route: it has been proposed that sulphur is released
into the gas phase mainly as H$_2$S, following also mantle sputtering,
and then H$_2$S is transformed into SO and SO$_2$ via neutral-neutral
reactions on timescales of $\sim 10^3$ yr 
\citep[e.g.][]{1997ApJ...481..396C}. Later on $\sim 3 \times
10^{4}$~yr, gaseous sulphur is expected to be in atomic sulphur.
Therefore, the constant SO$_{2}$/H$_{2}$CO ratio in E1, W1 and W2 may
tell us that these three shocks have similar ages, around $10^4$ yr if
models are correct, whereas E2 is a much older shock, $\geq 3 \times
10^4$ yr, and SO$_2$ and H$_2$CO both disappear because transformed
into S and complex O-bearing molecules respectively.

The overall emerging picture is that the combination of the variations
in the abundance of different molecules can in first instance give an
estimate of the age of the shock.  Table~\ref{ShockAge} summarizes the
situation. The times noted in the  Table~\ref{ShockAge} are really
modeled dependent and are meant to represent a  sequence.  The
simultaneous presence of SO, SO$_2$, H$_2$S, H$_2$CO, and SiO  would
mark relatively young shocks ($\leq 10^4$ yr), whereas the only
presence of SiO emission would point towards relatively older shocks
($\geq 3 \times 10^4$ yr).

\begin{table} 
\begin{tabular}{ccc} 
\hline
Time (yr)& Formation & Destruction \\
\hline 
\hline
0$^a$    & H$_2$S, H$_2$CO, SiH$_2$, SiH$_4$ & \\
$10^3$   & SO, SO$_2$, H$_2$S, SiO           & \\
$3 \times 10^4$   &  & SO$_2$, H$_2$CO\\
$10^5$   &   & SiO \\
\hline 
\end{tabular} 
\caption{Proposed schema of the formation and destruction of 
SO, SO$_2$, H$_2$S, SiO and H$_2$CO as function of time. 
The times are indicative and are meant to represent a sequence.
$^a$~Injection in the gas phase of the molecules sputtered 
from the grain mantles.}
\label{ShockAge} 
\end{table}

Finally, comparison with other molecular outflows can only be very
approximate, but still somewhat illustrative (Tab.~\ref{AbuRatio}).
The only molecular outflow at our knowledge in which SO, SO$_2$, SiO
and H$_2$CO abundances have been measured is the L1157 outflow
\citep{1997ApJ...487L..93B}.  Looking at Table \ref{AbuRatio}, L1157
B1 and B2 are rather similar to E2 with respect to the ratios
involving SiO, i.e. they are relatively enriched in SiO with respect
to the other shocks of L1689N and the molecular cloud. The likely
interpretation is that E2, B1 and B2 are all  relatively strong
shocks.  Yet, contrary to what happens in E2, H$_2$CO, SO and SO$_2$
are detected  in L1157 B1 and B2, which would point towards relatively
young shocks (Table \ref{ShockAge}).  Note though that SO seems to be
underabundant with respect to H$_2$CO and SiO in B1 and B2 when
compared not only to E2, but also to E1, W1 and W2.  At the same time,
SO$_2$ is overabundant.  Whether this is because of the stronger shock
in B1 and B2 that converts SO into SO$_2$ more quickly, or because the
shocks in L1157 have a different age of E1, W1 and W2, is impossible
to say at this stage.  A larger statistics on a larger number of
outflow system is necessary to better understand this point.

\subsection{IRAS16293}

\begin{figure}
\centering
\includegraphics[width=8cm]{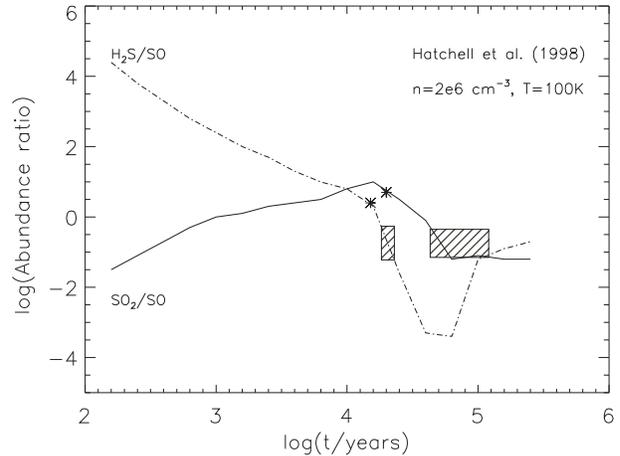}
\caption{Theoretical predictions of the
evolution of SO$_{2}$/SO and H$_{2}$S/SO ratios as a function
of time. The curves are adapted from  the \citet{1998A&A...338..713H}
model, that assumes a gas temperature of 100 K and a density of
$\rm 2 \times 10^{6} cm^{-3}$.
The grey squares represent the portion of the modelling curve in agreement with
observations of the hot core of IRAS16293 (this work).
Analogously, the stars mark the predicted values of the ratios
in agreement with what observed towards L1157-mm \citep{1997ApJ...487L..93B}.}
\label{x_tH}
\end{figure}

Table \ref{coresabund} quotes the absolute abundances of SO, SO$_{2}$,
H$_{2}$S, H$_{2}$CO and SiO in the hot core of IRAS16293 and in the
molecular cloud.  As evident from the table, the sulphuretted molecule
abundances exhibit a strong enhancement with respect to the molecular
cloud abundances.  The abundance of SO increases by a factor 200,
SO$_{2}$ by a  factor 1300 and H$_{2}$S by a factor 1700 \citep[see
also ][]{2002A&A...390.1001S}.

As we mentioned in the Introduction, the sulphur species are expected
to be useful to evaluate the age of hot cores.  We could not resist
the temptation to compare the measured abundance ratios with published
theoretical predictions.  In Fig. \ref{x_tH} we compare the H$_2$S/SO
and SO$_2$/SO ratios obtained in the IRAS16293 hot core with the
theoretical values predicted by chemical models.  The observed
H$_2$S/SO and SO$_2$/SO ratios are consistent with the theoretical
predictions by \citet{1998A&A...338..713H}, but differ substantially
with those by \citet{1997ApJ...481..396C}.  Both models use the
sulphur chemistry described in \citet{1997ApJ...481..396C} and assume
that the bulk of sulphur is stored in iced H$_2$S onto the grain
mantles. When the dust temperature exceeds 100 K, the grain mantles
evaporate releasing the H$_2$S into the gas phase, which is slowly
converted into SO and subsequently into SO$_2$ by neutral-neutral
reactions.  The only difference between the two models is that
\citeauthor{1998A&A...338..713H} use an updated rate for the reaction
of H$_2$S with atomic hydrogen. This difference substantially changes
the evolution of the abundances of H$_2$S, SO and SO$_2$ at times
larger than about $10^4$~yr.  Our observations support the
\citet{1998A&A...338..713H} model, as shown in Fig.~\ref{x_tH}.  The
agreement between observations and model predictions for an age of
$\sim 3\times 10^4$ yr is remarkable, as independent estimates of the
age of IRAS16293 converge towards the same value
\citep{2000A&A...355.1129C}.  On the same figure we also report the
values observed towards L1157-mm.  Based on these observations and on
the model predictions, L1157-mm has about the same age of IRAS16293.
We caution that the error bars are indeed large enough to make the
agreement between observed and predicted values just a
coincidence.  In this sense, this result may just mean an
encouragement to pursue this kind of studies on a large sample of low
mass protostars, where a statistical trend could be established.  We
have indeed started such a study on a sample of low mass protostars,
whose we have derived the physical structure (i.e. density and
temperature profiles) and formaldehyde abundance profiles 
\citep{2002A&A...395..573M}.

A direct comparison between the measured abundances in the  hot cores
quoted in Tab. \ref{coresabund} and that of IRAS16293 can only be done
with respect to Orion-KL, as in the other sources the abundances are
not corrected for the beam dilution 
\citep{1998A&A...338..713H,2003A&A...399..567B}.  With
respect to Orion-KL, the hot core of IRAS16293 is enriched in  SO,
SO$_2$ and H$_2$CO, whereas it is deficient in H$_2$S.  The most
plausible explanation is that the different abundances reflect a
different composition of the ices (from which H$_2$CO and  H$_2$S
evaporate), a fact now well documented by the study of deuterated
molecules in high and low mass protostars
\citep[e.g. ][]{2002P&SS...50.1267C}.  However, the different gas temperature
and density may also play a role.  Finally, the SO$_2$/SO and
H$_2$S/SO ratios of IRAS16293 differ by about a factor 100 and 10
respectively with respect to the relevant ratios measured in the high
mass protostars of the \citet{1998A&A...338..713H} sample. Again, the
difference may be attributed to both different  evolutionary stages
and different gas densities and temperatures.

Following the theoretical expectations that  gaseous sulphur in the
warm gas is mainly locked into SO, SO$_2$ and H$_2$S
\citep[e.g.][]{1993MNRAS.262..915P,1997ApJ...481..396C}, while silicon
is mainly locked into SiO \citep{1989A&A...222..205H}, the
(SO+SO$_2$+H$_2$S)/SiO ratio gives a measure of the gas phase
elemental sulphur over silicon  abundance ratio in the IRAS16293 hot
core. This value is about 300, to compare with the solar elemental
sulphur over silicon abundance ratio of  0.5. It is unlikely that we
overestimate the overall quantity of sulphur. Conversely, it could
have been underestimated because of a possible optical thickness of
the sulfuretted molecular lines and/or the presence of other important
sulphur-bearing molecular species not considered here. Thus, the
measured  (SO+SO$_2$+H$_2$S)/SiO ratio suggests that there is an
important deficiency of silicon in the warm gas of the IRAS16293 hot
core.  The likely  explanation is that silicon is mostly depleted into
the non-volatile, refractory cores grains,  whereas sulphur is
depleted mostly onto the volatile grain mantles.  This conclusion
supports the thesis by \citet{1999MNRAS.306..691R} that sulphur and
silicon follow different routes of depletion, and specifically that
sulphur is depleted at the time of mantle formation.  Another
possibility to explain this deficiency of silicon is that  silicon is
mainly in another form than SiO in the warm gas such as  Si, Si$^{+}$
or SiO$_{2}$.  The atomic silicon is difficult to observe, since it
has its ground transition at 129~$\mu$m, i.e. a wavelength range
obscured by the atmosphere. Observations obtained with the  {\it Long
Wavelength Spectrometer} on board ISO did not detect any signal larger
than about $5 \times 10^{-13}$~erg~s$^{-1}$~cm$^{-2}$
\citep{1998A&A...331L..17C}, corresponding to an upper limit for the
Si column density of $6 \times 10^{17}$~cm$^{-2}$, and an abundance
lower than $8 \times 10^{-6}$,  i.e. about 500 times more than the SiO
abundance found in IRAS16293.  For Si$^{+}$, it is unlikely that the
UV field in the observed regions is  strong enough to ionize the
silicon.

Finally, the total abundance of SO, SO$_{2}$ and H$_{2}$S in the
IRAS16293 hot core is $2.8 \times 10^{-6}$, namely more than ten times
lower than the corresponding  solar abundance $3.4 \times
10^{-5}$. Even adding up the OCS, whose abundance is about
that of SO \citep{2002A&A...390.1001S}, 
the overall sulphur abundance is still
low. Unfortunately \citep{2002A&A...390.1001S} could not estimate
the CS abundance in the hot core of IRAS16293, but CS is unlikely
to be the main reservoir of sulphur.
The other possible reservoir of sulphur, the atomic sulphur
\citep{1997ApJ...481..396C}, is extremely difficult to observe,
because the ground state transition is at 25~$\mu$m, i.e. in
a wavelength range obscured by  the atmosphere. 

\section{Conclusion}

We have presented a quantitative observational study of the most
important S-bearing molecules, namely SO, SO$_{2}$ and H$_{2}$S, in
the region of L1689N.  We derived the column density of these
molecules plus SiO and H$_2$CO molecules in six regions of L1689N: the
cloud, the young protostar IRAS16293, and four shocked regions.

We found that SiO is the molecule that shows the largest  abundance
variations in the shocked regions, whereas S-bearing molecules show
more moderate variations.  Remarkably, the region of the brightest SiO
emission in L1689N, namely E2, is undetected in SO$_2$, H$_2$S and
H$_2$CO and only marginally  detected in SO. We argued that this is
possibly due to the relatively old age ($\geq 3 \times 10^4$ yr)  of
this shock.

In the other weaker SiO shocks, SO$_2$ is enhanced with respect to SO,  in
agreement with theoretical  expectations that predict the conversion
of the gaseous sulphur mostly into SO$_2$ on timescales of $\sim 10^3$
yr.  In the same regions, the SO$_2$/H$_2$CO ratio is of order of
unity.  We argued that this may point to relatively young shocks
($\sim 10^4$ yr), where SO$_2$ has already formed and H$_2$CO has not
yet destroyed.

Putting together the observed combinations of the SO, SO$_2$, H$_2$CO
and SiO ratios, we proposed a schema in which the different molecular
ratios correspond to different ages of the shocks.

Finally, we found that SO, SO$_2$ and H$_2$S have significant
abundance jumps (200, 1300 and 1700 respectively) in the inner hot
core of IRAS16293.  We compared the measured abundances with
theoretical models and discussed the derived protostar age. However, we 
cautioned that a more detailed study is necessary to draw reliable conclusions. 
  The hot
core of IRAS16293 seems to be enriched in SO, SO$_2$ and H$_2$CO with
respect to Orion-KL, probably because of a different initial
composition of the ices in the two sources.  Comparing the
SO+SO$_2$+H$_2$S/SiO ratio in the hot core of IRAS16293,  we found
that silicon is largely deficient in the warm gas (by a factor $\sim
600$), supporting the thesis that silicon is depleted into the grain
refractory cores whereas sulphur is depleted into the grain volatile
mantles.  Nonetheless, sulphur in the IRAS16293 warm gas is also
deficient.

\begin{acknowledgements}
We thank the IRAM and SEST staff in Pico Veleta and La Silla for their
assistance with the observations, and the IRAM and ESO Program
Committee for their award of observing time. We would like to thank G. 
Fuller, the referee, for useful comments.We are grateful to
A.G.G.M. Tielens and M. Walmsley, for helpful discussions on sulphur and 
silicon chemistry. V. Wakelam wishes to thank F. Herpin for his help on data
reduction. 
\end{acknowledgements}



\end{document}